\documentclass[fleqn,twoside]{article}
\usepackage{espcrc2}
\usepackage{epsfig}

\usepackage{graphicx}
\usepackage[figuresright]{rotating}

\newcommand{\AmS}{{\protect\the\textfont2
  A\kern-.1667em\lower.5ex\hbox{M}\kern-.125emS}}

\title{Charged Current Quasi-Elastic Interactions at MiniBooNE Confront Cross Section Monte Carlos}

\author{J. Monroe\address[MCSD]{Columbia University Department of Physics,\\ 538 W. 120th Street, New York, NY, 10027, USA} for the MiniBooNE Collaboration\thanks{MiniBooNE gratefully acknowledges the support it receives from the Department of Energy and from the National Science Foundation.  The presenter of this paper was supported by NSF grant PHY-98-13383.}}

\begin{document}

\begin{abstract}
Neutrino oscillations have been established in solar and atmospheric neutrinos, but a third signal from the LSND experiment is incompatible with three Standard Model neutrinos.  The MiniBooNE experiment can confirm or refute the LSND oscillation signal with $1 \times 10^{21}$ protons on target.  While working towards the oscillation result, MiniBooNE will accumulate more than $1 \times 10^6$ neutrino interactions in the 0 to 2 $GeV$ range which will greatly increase the world's knowledge of neutrino cross sections in this energy regime.  Preliminary results on the MiniBooNE $\nu_{\mu}$ charged current quasi-elastic analysis are presented and compared to the NUANCE, NEUT, and NEUGEN cross section Monte Carlos.
\vspace{1pc}
\end{abstract}

\maketitle

\section{Introduction}

Neutrino masses have been conclusively established via positive neutrino oscillation measurements of solar and atmospheric neutrinos \cite{solar}, \cite{atmospheric}.  A third and unconfirmed positive result comes from the LSND experiment, a short baseline accelerator $\overline{\nu}_{\mu} \rightarrow \overline{\nu}_e$ oscillation search \cite{lsndfinal}.  The MiniBooNE experiment at Fermi National Accelerator Laboratory will confirm or refute the LSND result with higher statistics and different sources of systematic error.  

MiniBooNE will accumulate more than $1 \times 10^6$ neutrino interactions on a $CH_2$ target.  There is a dearth of neutrino cross section data on heavy targets in the MiniBooNE energy range, $0 < E_{\nu} < 2$ $GeV$.  Further, this is a complex region theoretically since both charged current quasi-elastic (CCQE) and resonance processes contribute in roughly equal proportions.  

The description of this region differs significantly among neutrino cross section Monte Carlos.  Collaboration between the authors of the NUANCE\cite{nuance}, NEUT\cite{neut}, and NEUGEN\cite{neugen} Monte Carlos and the MiniBooNE cross sections working group has made detailed comparisons of CCQE event rates and kinematics possible for the first time.  Confronting the predictions of these cross section Monte Carlos with the MiniBooNE CCQE data set provides valuable input to the development of simulations of this complex regime.  Preliminary indications point to deficiencies in the modeling of low $Q^2$ interactions common to all cross section Monte Carlos considered here.    

\section{MiniBooNE Overview \label{overview}}

\begin{figure}[ht]
\vspace{-0.5cm}
\epsfxsize=2.8in\epsfbox{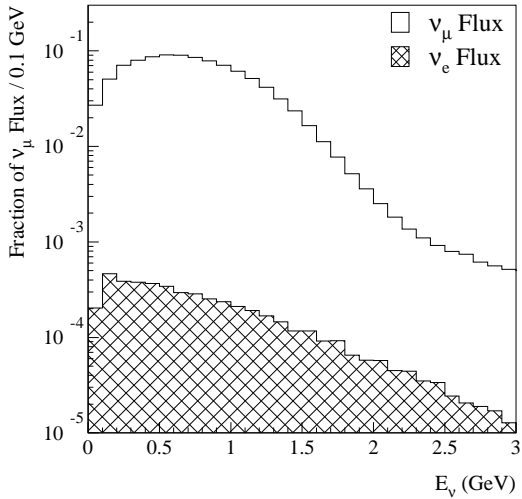}
\vspace{-1.0cm}
\caption{Preliminary MiniBooNE neutrino flux Monte Carlo prediction $vs$. $E_{\nu}$ ($GeV$).\label{fig:mboone_flux}}
\vspace{-1.cm}
\end{figure}

The MiniBooNE neutrino beam is produced from 8.89 $GeV/c$ protons incident on a thick beryllium target located inside a magnetic focusing horn.  Typical operating conditions are $4 \times 10^{12}$ protons per pulse, at 3 - 4 $Hz$ with a beam uptime of $\sim 88\%$.  The beam spill duration is 1.6 $\mu s$.  The focusing horn increases the neutrino flux at the detector by a factor of $\sim$5, and can operate in both positive and negative polarities for $\nu$ and $\overline{\nu}$ running; currently positive sign mesons are selected.  Mesons produced in the target decay in a 50 $m$ long decay pipe.  The neutrino beam is primarily composed of $\nu_{\mu}$ from $\pi^+$ decay with a 0.5\% contamination from $\nu_e$.  The mean neutrino energy is $\sim 700$ $MeV$; the neutrino flux is shown in figure \ref{fig:mboone_flux}.

The MiniBooNE detector is a 6.1 $m$ radius sphere filled with mineral oil ($CH_2$).  There are 1280 inward-facing ``tank'' PMTs, and 240 outward-facing ``veto'' PMTs.  Particle identification depends upon both prompt Cerenkov and time-delayed scintillation light.  Neutrino induced events are identified by requiring that the event occur within the beam spill, have fewer than 6 veto PMT hits, and have greater than 200 tank PMT hits.  With these simple cuts the cosmic ray background is reduced to less than 0.1\% of the beam-induced neutrino signal\footnote{The cosmic ray rejection is demonstrated in reference \cite{jen}.}.  A fiducial volume cut at $R < 5 \ m$ is also typically required to ensure good energy reconstruction.  

\section{$\nu_{\mu}$ Event Rate Prediction}
The event rate prediction is based on the product of neutrino flux and cross section.  The MiniBooNE neutrino flux is primarily produced from $\pi^+$ decay in flight.  Therefore a detailed understanding of the $\pi^+$ production in $p \ - Be$ collisions is necessary.  The neutrino cross section at MiniBooNE energies has contributions from a number of different processes \cite{lipari}.  Disentangling the various contributions to the total cross section is of theoretical interest in this energy regime, and is important to oscillation measurements.

\subsection{Flux Prediction}
At MiniBooNE, the relevant ranges of $\pi^+$ production momenta and angles are $1 < p_{\pi} < 4$ $GeV/c$ and $0 < \theta_{\pi} < 0.2$ radians respectively.  There are no existing measurements at 8.89 $GeV/c$ proton momentum.  However there is a limited amount of relevant production data from past experiments \cite{piprodexpts} at 10, 12, and 19 $GeV/c$.  To address the paucity of production information, MiniBooNE collaborators have analyzed data at 6, 12, and 17 $GeV/c$ from the BNL E910 experiment \cite{e910}.  In addition, 20 million triggers were collected with a replica MiniBooNE target and an 8.89 $GeV/c$ proton beam at the CERN HARP experiment \cite{harp}.  The analysis of this data is currently in progress, and the resulting $p \ Be \ \rightarrow \ \pi^+ \ X$ cross section measurement will be used for the final MiniBooNE flux prediction.  Currently, the neutrino flux is modeled with a GEANT4 based Monte Carlo \cite{G4} and an external parameterization of the $p \ Be \ \rightarrow \ \pi^+ \ X$ cross section.  The parametrization comes from a global fit of existing production data in the range $10 < p_{proton} < 17 GeV/c$ to the Sanford-Wang model \cite{SW}.
  
\subsection{Cross Section Prediction}
The NUANCE Monte Carlo is used to predict the neutrino interaction cross sections.  At MiniBooNE neutrino energies the cross section has contributions from charged current quasi-elastic scattering (39\% of the total event rate), charged current resonance production (25\%), neutral current elastic scattering (7\%), and neutral current $\pi^0$ production (7\%).  For the $\nu_{\mu} \rightarrow \nu_e$ oscillation analysis, the most important processes are CCQE scattering which affords a precise measurement of the neutrino energy, NC $\pi^0$ production which is a large background to a $\nu_e$ signal \cite{jen}, and NC elastic scattering which can be used to study the optical properties of the detector and nuclear recoil.

\section{CCQE Events \label{CCQEsection}}
\begin{figure}[ht]
\vspace{-0.8cm}
\epsfxsize=2.8in\epsfbox{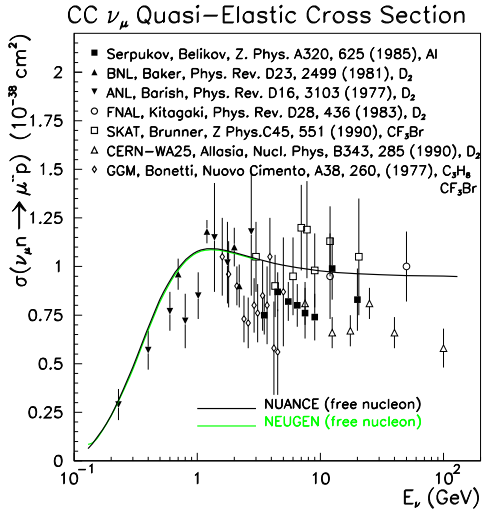}
\vspace{-1.0cm}
\caption{Low energy CCQE cross section measurements $vs$. $E_{\nu}$ ($GeV$) \cite{sam}.\label{CCQExsecsum}}
\vspace{-0.5cm}
\end{figure}

Charged current quasi-elastic interactions are fairly well measured in the MiniBooNE energy range on light targets.  However, the cross section uncertainty on this process is an important error contribution to oscillation searches, particularly for heavy target experiments like MiniBooNE.  A summary of the existing data is shown in figure \ref{CCQExsecsum}.  

Quasi-elastic kinematics enable a precise determination of the neutrino energy in $\nu_{\mu} (\nu_e) \ n \rightarrow \ \mu^- (e^-) \ p$ interactions.  Neglecting corrections for the motion of the target nucleon, the neutrino energy can be calculated from the measured energy and angle of the final state muon:
\begin{equation}
E_{\nu}^{QE} \ = \ \frac{1}{2} \frac{2 M_{p} E_{\mu} - m_{\mu}^2}{M_p - E_{\mu} + \sqrt{(E_{\mu}^2 - m_{\mu}^2)} \cos \theta_{\mu}}
\end{equation}
where $M_p$ is the proton mass, $m_{\mu}$ is the muon mass, $E_{\mu}$ is the muon energy, and $\theta_{\mu}$ is the muon angle with respect to the beam direction.  The energy resolution achievable by MiniBooNE assuming no non-CCQE background is $\sim$10\% at $E_{\nu}$ = 1 $GeV$.

\section{Comparison of Monte Carlos}

Charged current quasi-elastic scattering is a simple process and therefore enables a straight forward comparison of cross section Monte Carlos.  The NUANCE version 2 and version 3, NEUT, and NEUGEN Monte Carlos have been compared using the MiniBooNE flux, detector Monte Carlo, and reconstruction.  This is the first such ``apples-to-apples'' comparison for CCQE interactions among cross section Monte Carlos.
  
\subsection{Theoretical Inputs}
The NUANCE, NEUT, and NEUGEN Monte Carlos have common theoretical inputs such as the LLewellyn-Smith free nucleon quasi-elastic cross section \cite{LlewellynSmith}, the Rein-Sehgal resonance cross section model \cite{Rein-Sehgal_res}, and the standard deep inelastic scattering formula for high $Q^2$ \cite{DIS}.  However there are non-trivial differences as well.  These include the implementation of the Fermi gas model for quasi-elastic interactions, the method for joining the resonance and deep inelastic scattering regions, and the treatment of final state interactions.

The following is a brief summary of the salient theoretical inputs for these Monte Carlos.  NUANCE version 2 uses dipole form factors, the Smith-Moniz Fermi gas model \cite{SmithMoniz}, and $m_A$ = 1.0 $GeV/c$$^2$.  NUANCE version 3 uses non-dipole form factors \cite{nondipole}, a new $\pi$ absorption model tuned on $\pi$ data \cite{hawker}, and $m_A$ = 1.03 $GeV/c$$^2$.  NEUT uses dipole form factors, the Smith-Moniz Fermi gas model, and $m_A$ = 1.1 $GeV/c$$^2$.  NEUGEN uses dipole form factors, a $\pi$ absorption model tuned on $\nu$ data, the Bodek-Ritchie modified Fermi gas model \cite{BodekRitchie}, no nucleon re-scattering, and $m_A$ = 1.032 $GeV/c$$^2$.

\subsection{Normalization}
The predicted CCQE interaction rate is very similar between the NUANCE, NEUT, and NEUGEN Monte Carlos.  The CCQE percentage of the total number of events is 38.1\% (NUANCE version 2), 39.8\% (NUANCE version 3), 38.0\% (NEUT), and 38.0\% (NEUGEN).

\subsection{Kinematics}
The kinematic distributions for CCQE events considered here are reconstructed visible energy, reconstructed angle with respect to the neutrino beam direction, reconstructed quasi-elastic neutrino energy, and reconstructed $Q^2$.
\begin{figure*}
\vspace{-0.5cm}
\hspace{1.5cm}
\epsfxsize=4.8in\epsfbox{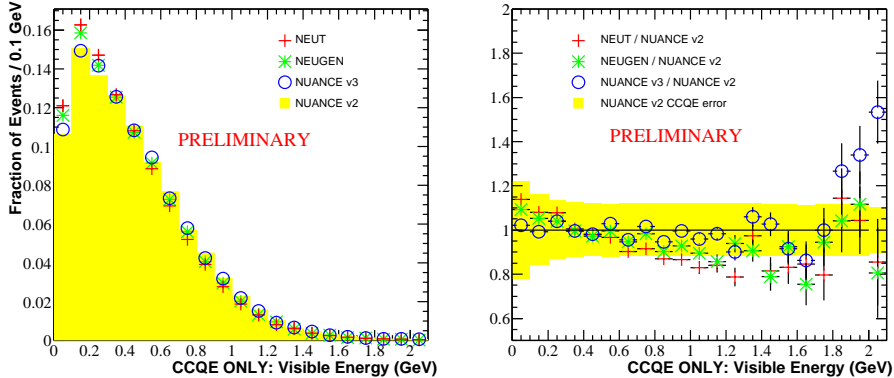}
\vspace{-0.5in}
\caption{Left: reconstructed visible energy ($GeV$) of Monte Carlo $\nu_{\mu}$ CCQE events, normalized to unit area.  Right: ratio of Monte Carlo predictions to NUANCE version 2 $vs$. reconstructed visible energy ($GeV$), shown with NUANCE version 2 CCQE cross section error band about 1.0. \label{fig:CCQE_evis}}
\end{figure*}

The visible energy is calculated from prompt light in the reconstruction, and is approximately equivalent to the muon kinetic energy.  A comparison of the visible energy distributions from the various cross section Monte Carlos is shown in figure \ref{fig:CCQE_evis} (left), all normalized to unit area.  NUANCE version 2 has a slightly harder spectrum, as can be seen in the ratio of the various Monte Carlos to the NUANCE version 2 prediction (figure \ref{fig:CCQE_evis}, right).  The band about 1.0 indicates the CCQE cross section error currently assumed by MiniBooNE.  

\begin{figure*}
\vspace{-0.5cm}
\hspace{1.5cm}
\epsfxsize=4.8in\epsfbox{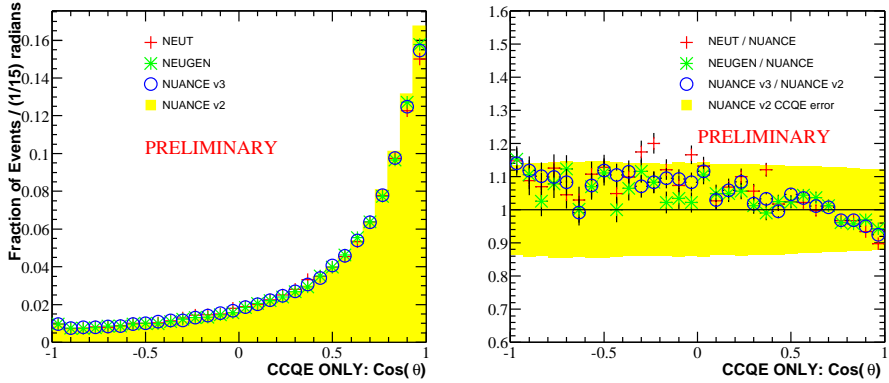}
\vspace{-0.5in}
\caption{Left: reconstructed $\cos(\theta)$ of Monte Carlo $\nu_{\mu}$ CCQE events, normalized to unit area.  Right: ratio of Monte Carlo predictions to NUANCE version 2 $vs$. reconstructed $\cos(\theta)$, shown with NUANCE version 2 CCQE cross section error band about 1.0. \label{fig:CCQE_cost}}
\end{figure*}

The reconstructed angle of the final state muon with respect to the neutrino beam direction is shown in figure \ref{fig:CCQE_cost}.  There are significant differences between Monte Carlos in this distribution, particularly at high $\cos(\theta)$ where NUANCE version 3, NEUT, and NEUGEN are all $\sim$5 - 10\% lower than NUANCE version 2.  

\begin{figure*}
\vspace{-0.5cm}
\hspace{1.5cm}
\epsfxsize=4.8in\epsfbox{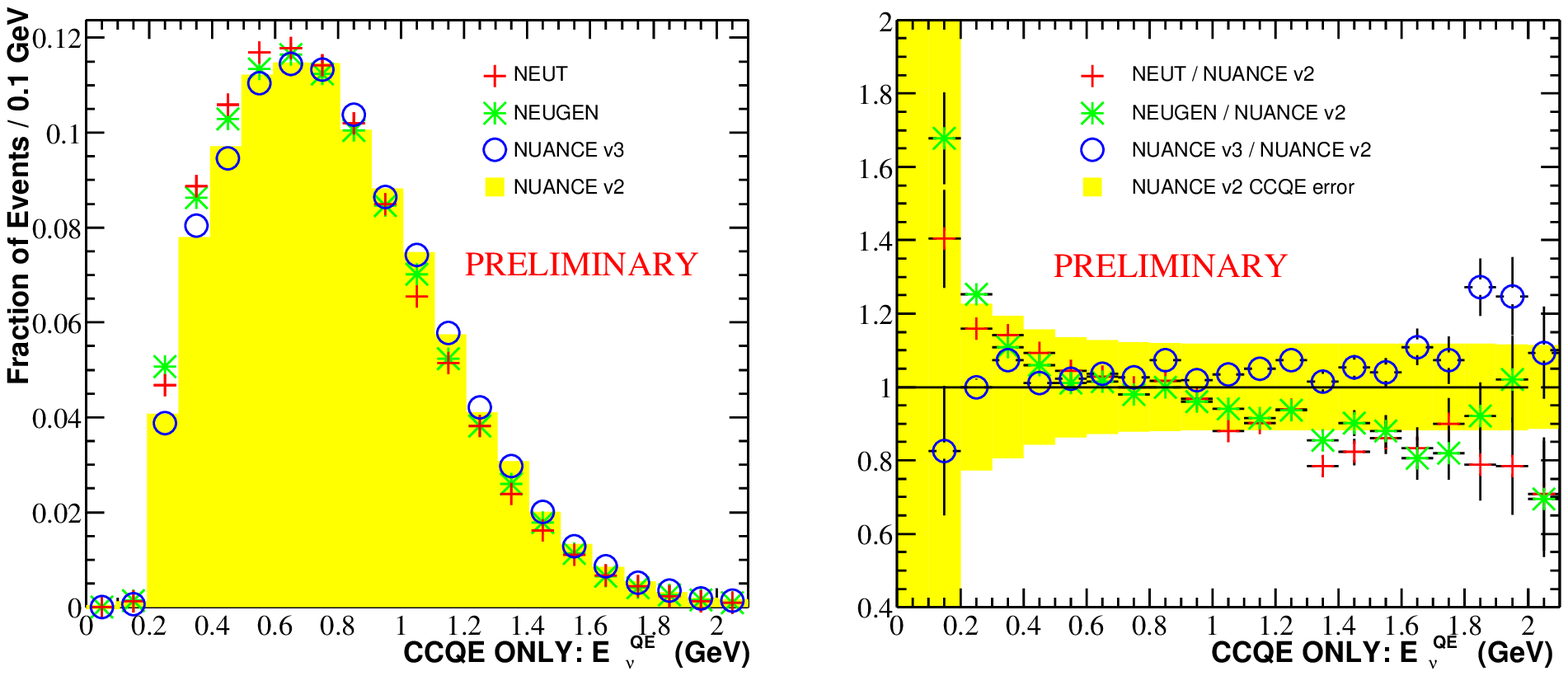}
\vspace{-0.5in}
\caption{Left: reconstructed $E_{\nu}^{QE}$ ($GeV$) of Monte Carlo $\nu_{\mu}$ CCQE events, normalized to unit area. Right: ratio of Monte Carlo predictions to NUANCE version 2 $vs$. reconstructed $E_{\nu}^{QE}$ ($GeV$), shown with NUANCE version 2 CCQE cross section error band about 1.0. \label{fig:CCQE_enuqe}}
\end{figure*}

The neutrino energy, calculated from the reconstructed muon energy and angle, is shown in figure \ref{fig:CCQE_enuqe}.  For $E_{\nu}^{QE}$, both NEUT and NEUGEN exhibit softer spectra than NUANCE versions 2 and 3.  

\begin{figure*}
\vspace{-0.5cm}
\hspace{1.5cm}
\epsfxsize=4.8in\epsfbox{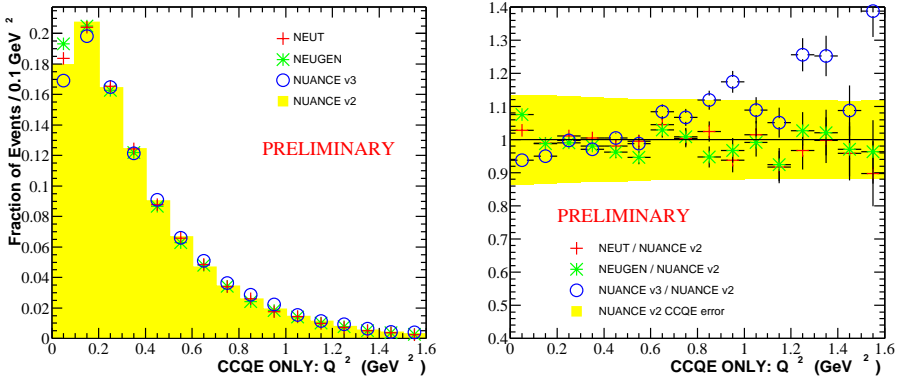}
\vspace{-0.5in}
\caption{Left: reconstructed $Q^2$ ($GeV$$^2$) of Monte Carlo $\nu_{\mu}$ CCQE events, normalized to unit area. Right: ratio of Monte Carlo predictions to NUANCE version 2 $vs$. reconstructed $Q^2$, ($GeV^2$) shown with NUANCE version 2 CCQE cross section error band about 1.0.  \label{fig:CCQE_q2}}
\end{figure*}

The reconstructed $Q^2$ distributions are shown in figure \ref{fig:CCQE_q2}.  The differences of $\sim \pm$10\% at $Q^2 \ = \ 0$ between Monte Carlos are consistent with the comparisons of the $\cos(\theta)$ and visible energy distributions.

In general, it is striking that the NUANCE, NEUT, and NEUGEN Monte Carlos all agree within errors given the differences in their theoretical inputs.  The largest disagreements occur at high $\cos(\theta)$ and low $Q^2$.  It may be significant that this is the regime where the theoretically ``messy'' components of CCQE interactions, Pauli blocking and final state interactions, are most important.

\section{Comparison with MiniBooNE Data}

MiniBooNE data will eventually be used to measure the $\nu_{\mu}$ CCQE cross section, and even at this preliminary stage can be used to confront Monte Carlo predictions for event kinematics.  To compare the cross section Monte Carlo predictions with MiniBooNE data, the same event selection cuts are applied to isolate CCQE events.  The kinematic distributions discussed in the previous section are then compared with the MiniBooNE CCQE data set.

\subsection{CCQE Event Selection}
The CCQE selection criteria require that the event pass the cosmic background and fiducial volume cuts described in section \ref{overview}, and that the event topology be consistent with a single muon-induced Cerenkov ring.  The efficiency of the event selection is $\sim$30\%.  For events generated within the 5 $m$ fiducial volume the efficiency is $\sim$55\%.  The background is almost entirely due to charged current single pion production events, $\nu_{\mu} \ p \ \rightarrow \ \mu^- \ p \ \pi^+$ where the $\pi^+$ is produced below its Cerenkov threshold.  The current MiniBooNE $\nu_{\mu}$ CCQE data set comprises $\sim$60,000 events.  

The event selection performance varies slightly for the different Monte Carlos.  A comparison of the CCQE interaction rate before and after the event selection cuts is shown in table 1.  The resulting CCQE purity and background contamination is shown in table 2.
\begin{table}[htb]
\vspace{-0.5cm}
\caption{Normalization comparison of Monte Carlos before and after CCQE event selection cuts.  The ratio shown is the number of events with respect to NUANCE version 2.}
\vspace{0.2cm}
\begin{tabular}{@{}llll}
\hline
\emph{Monte} & \emph{ratio be-} & \emph{Efficiency} & \emph{ratio af-} \\
\emph{Carlo} & \emph{fore cuts} & \emph{of cuts \%} & \emph{ter cuts} \\
\hline
NUANCE v2 & 1.0  & 24.8 & 1.0 \\
NUANCE v3 & 1.03 & 24.8 & 1.05 \\
NEUT      & 0.98 & 24.5 & 1.07 \\
NEUGEN    & 0.98 & 25.2 & 1.0 \\
\hline
\end{tabular}\\[2pt]
\end{table}
\begin{table}[htb]
\vspace{-1.5cm}
\caption{CCQE purity comparison of Monte Carlos after event selection cuts.  The dominant background is resonant single $\pi$ production, $\nu_{\mu} \ p \rightarrow \mu^- \ p \ \pi^+$.}
\vspace{0.2cm}
\begin{tabular}{@{}llll}
\hline
\emph{Monte} & \emph{CCQE} & \emph{Resonant} & \emph{\% other} \\
\emph{Carlo} & \emph{\%}        & \emph{1 $\pi$ \%} & \emph{non-CCQE} \\
\hline
NUANCE v2 & 83 & 14 & 3 \\
NUANCE v3 & 80 & 16 & 4 \\
NEUT      & 78 & 13 & 9 \\
NEUGEN    & 80 & 16 & 4 \\
\hline
\end{tabular}\\[2pt]
\vspace{-1.cm}
\end{table}

\subsection{Kinematics Comparison}

\begin{figure}
\vspace{-0.3cm}
\epsfxsize=2.4in\epsfbox{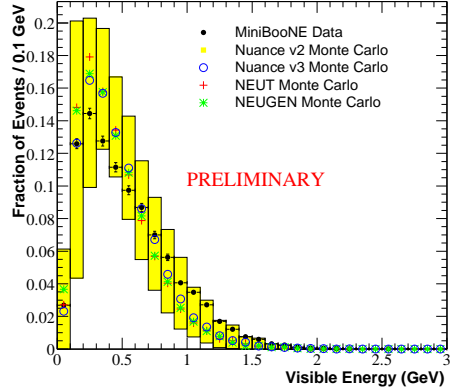}
\vspace{-0.5in}
\caption{Reconstructed visible energy ($GeV$) of $\nu_{\mu}$ CCQE Monte Carlo and data, unit area normalized. \label{fig:evis_wdata}}
\vspace{-0.5cm}
\end{figure}

\begin{figure}
\epsfxsize=2.4in\epsfbox{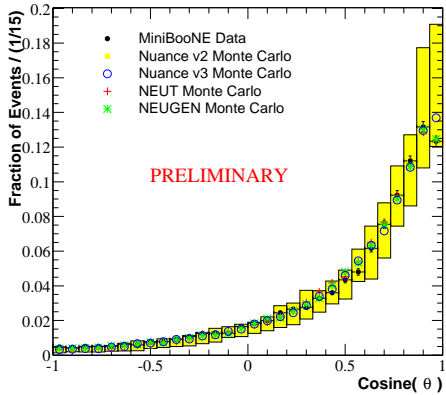}
\vspace{-0.5in}
\caption{Reconstructed $\cos(\theta)$ of $\nu_{\mu}$ CCQE Monte Carlo and data, unit area normalized. \label{fig:cost_wdata}}
\vspace{-0.5cm}
\end{figure}

\begin{figure}
\epsfxsize=2.4in\epsfbox{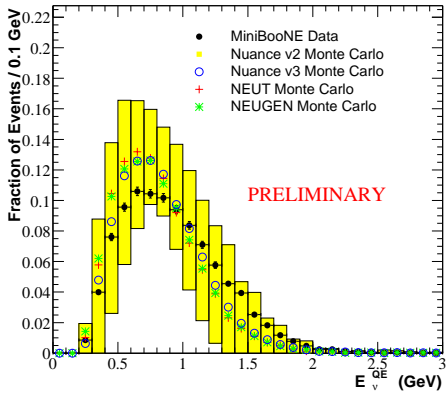}
\vspace{-0.5in}
\caption{Reconstructed $E_{\nu}^{QE}$ ($GeV$) of $\nu_{\mu}$ CCQE Monte Carlo and data, unit area normalized. \label{fig:enuqe_wdata}}
\vspace{-0.5cm}
\end{figure}

\begin{figure}
\epsfxsize=2.4in\epsfbox{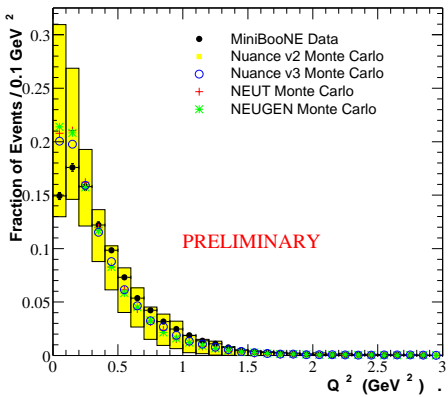}
\vspace{-0.5in}
\caption{Reconstructed $Q^2$ ($GeV$$^2$) of $\nu_{\mu}$ CCQE Monte Carlo and data, unit area normalized. \label{fig:q2_wdata}}
\vspace{-0.5cm}
\end{figure}

The visible energy, $\cos(\theta)$, $E_{\nu}^{QE}$, and $Q^2$ distributions after CCQE selection cuts are shown in figures \ref{fig:evis_wdata}, \ref{fig:cost_wdata}, \ref{fig:enuqe_wdata}, and \ref{fig:q2_wdata} respectively.
The bands on the NUANCE version 2 Monte Carlo prediction indicate uncertainties from flux, CCQE cross section, and detector optical model variations.  The data are shown with statistical errors.  The NUANCE version 3, NEUT, and NEUGEN Monte Carlo predictions are overlaid.  All distributions are normalized to unit area.

The shape of the data in the visible energy distribution is significantly harder than the predictions of the various cross section Monte Carlos, although the data are within the current error band.  The agreement between Monte Carlos and data is good for all $\cos(\theta)$ bins other than $\cos(\theta) \ \sim$ 1 where the NUANCE version 3, NEUT, and NEUGEN Monte Carlos agree with the data much better than NUANCE version 2.  The data exhibit a somewhat harder spectrum in $E_{\nu}^{QE}$ than the Monte Carlo predictions.  The most striking difference between data and Monte Carlo occurs near $Q^2 \ \sim$ 0 where the data are far lower than the predictions.  

It is interesting that although the $\cos(\theta)$ distributions agree reasonably well for the data and some Monte Carlos, no Monte Carlo reproduces the low $Q^2$ behavior of the data.  This may suggest a model deficiency common to all Monte Carlos considered here.

\section{Conclusions}

For the first time, ``apples-to-apples'' comparisons have been made between cross section Monte Carlos.  The NUANCE, NEUT, and NEUGEN simulations agree surprisingly well with each other for CCQE interaction rates and kinematics, given the differences in their theoretical inputs.  There is disagreement at the 10\% level  at very low $Q^2$ between the MiniBooNE CCQE data set and all Monte Carlos considered here.  In this regime, nuclear effects become most important.  The differences between Monte Carlos and the MiniBooNE data may indicate a need for more sophisticated models of low $Q^2$ neutrino events.


\begin{thebibliography}{9}
\bibitem{solar} For example: Q.~R.~Ahmad {\it et al.}, Phys.\ Rev.\ Lett.\  {\bf 89}, 011302 (2002). 

\bibitem{atmospheric} For example: M.~H.~Ahn {\it et al.}, Phys.~Rev.~Lett.~{\bf 90}, 041801 (2003).

\bibitem{lsndfinal} 
A.~Aguilar {\it et al.}, Phys. Rev. {\bf D64}, 112007 (2001).

\bibitem{nuance} 
D. Casper, Nucl. Phys. B Proc. Suppl. {\bf{112}}, 161 (2002).

\bibitem{neut}
Y. Hayato, Nucl. Phys. B Proc. Suppl. {\bf{112}}, 171 (2002).

\bibitem{neugen}
H. Gallagher, Nucl. Phys. B Proc. Suppl. {\bf{112}}, 188 (2002).

\bibitem{jen} J.~L.~Raaf, proceedings of this conference.

\bibitem{lipari} 
P. Lipari, Nucl. Phys. B Proc. Suppl. {\bf{112}}, 274 (2002).

\bibitem{piprodexpts} 
For example:
Cho, {\it et. al.}, Phys. Rev. {\bf D4}, 1967 (1971).

\bibitem{e910} I.~Chemakin {\it et al.}, Phys. Rev. {\bf C65}, 024904 (2002).

\bibitem{harp} The HARP collaboration: \\ \texttt{http://harp.web.cern.ch/harp/}.

\bibitem{G4} The Geant4 collaboration: \\ \texttt{http://geant4.web.cern.ch/geant4/}.

\bibitem{SW} J.~R.~Sanford and C.~L.~Wang, BNL AGS internal report \# BNL11299 and \# BNL11479, (1967).

\bibitem{sam} 
G. P. Zeller, hep$-$ex/0312061 (2002).

\bibitem{LlewellynSmith}
C.H. Llewellyn Smith, Phys. Rep. {\bf 3C}, 261 (1972).

\bibitem{Rein-Sehgal_res}
D. Rein and L.M. Sehgal, Annals of Physics {\bf 133}, 79 (1981).

\bibitem{DIS}
C. H. Albright and C. Jarlskog, Nucl. Phys. {\bf B84}, 467 (1975).


\bibitem{SmithMoniz}
R. A. Smith and E. J. Moniz, Nucl. Phys. {\bf B43}, 605 (1972).

\bibitem{nondipole}
P. E. Bosted, Phys. Rev. {\bf C51}, 409 (1995).

\bibitem{hawker} E. Hawker, proceedings of this conference.

\bibitem{BodekRitchie}
A. Bodek and J. L. Ritchie, Phys. Rev. {\bf D24}, 1400 (1981).


\end{thebibliography}
\end{document}